\documentclass[journal,onecolumn,12pt]{IEEEtran}
\usepackage[utf8]{inputenc}
\usepackage{amsmath,amssymb}
\usepackage{graphicx}                                   
\usepackage{url}   
\usepackage{cite} 
\usepackage[tight,footnotesize]{subfigure}  
\usepackage{amsthm}
\usepackage{array,multirow}
\usepackage{psfrag}
\usepackage{color, colortbl}  
\definecolor{LightCyan}{rgb}{0.88,1,1}
\definecolor{Gray}{rgb}{0.82,0.82,0.82}

\newcommand{\xbf}{\mathbf{x}}

\newcommand{\wbf}{\mathbf{w}}

\newcommand{\fbf}{\mathbf{f}}
\newcommand{\Fbf}{\mathbf{F}}

\newcommand{\gammabar}{\overline{\gamma}}

\theoremstyle{plain}

\ifCLASSINFOpdf
\else
\fi
\begin{document}
%
\title{Low-Complexity Set-Membership Normalized LMS Algorithm for Sparse System Modeling}
%
%
%

\author{Javad Sharafi and Mohsen Mehrali-Varjani~\IEEEmembership{}
\thanks{(J. Sharafi), Imam Ali University, Tehran, Iran. E-mail address: javadsharafi@grad.kashanu.ac.ir.}
\thanks{(M. Mehrali-Varjani), Imam Ali University, Tehran, Iran. E-mail address: Mohsenmehrali@yahoo.com.}
}

%
%

\maketitle

\begin{abstract}
In this work, we propose two low-complexity set-membership normalized least-mean-square (LCSM-NLMS1 and LCSM-NLMS2) algorithms to exploit the sparsity of an unknown system. 
For this purpose, in the LCSM-NLMS1 algorithm, we employ a function called the discard function to the adaptive coefficients 
in order to neglect the coefficients close to zero in the update process. 
Moreover, in the LCSM-NLMS2 algorithm, to decrease the overall number of computations needed even further, we substitute small coefficients with zero. 
Numerical results present similar performance of these algorithms when comparing them with 
some state-of-the-art sparsity-aware algorithms, whereas the proposed algorithms 
need lower computational cost. 
\end{abstract}


%
\IEEEpeerreviewmaketitle

\section{Introduction}
%
%
%
%
Adaptive filtering has applications in many areas such as communications, 
control, radar, acoustics, and speech processing. 
Nowadays, sparsity is an ubiquitous characteristic in signal or system parameters. 
Unfortunately, traditional adaptive filtering algorithms, such as the least-mean square (LMS) and the normalized LMS (NLMS) algorithms, do not exploit the sparsity in the signal or system models to improve the learning performance. 

We know that that by exploiting signal sparsity, we can remarkably improve the convergence rate and/or the steady-state performance of the learning process. 
Therefore, many improvement in the classical algorithms were introduced to exploit sparsity. 
A well-known family of algorithms to exploit sparsity is the family of proportionate algorithms~\cite{Duttweiler_PNLMS_tsap2000,Benesty_IPNLMS_icassp2002,Mula8627952,Diniz_sm_pap_jasmp2007,Salman8404502}. 
Another famous strategy to exploit sparsity is obtained by including a penalty function, such as the $l_1$-norm and the $l_0$-norm, to the cost function of traditional algorithms\cite{Theodoridis_l1ball_tsp2011,Gu_l0_LMS_SPletter2009,Yazdanpanah_globalsip,Yazdanpanah_RLS,Theodoridis_distributedl1ball_tsp2012}.
Note that in these methods something is added  
to the conventional algorithms, thus they increase computational cost of the algorithms.

In this work, to exploit systems sparsity, we employ a different approach from adding some features to the algorithms. 
Indeed, we ignore coefficients close to zero; thus, we can decrease the computational resources.
It is good to mention that a sparse impulse response of a system contains a few coefficients 
with high energy, and most of the coefficients are close to zero. 
Therefore, by using some knowledge about the uncertainty of small coefficients, 
we may substitute the coefficients smaller than assumed uncertainty with 
zero to reduce computational costs. In other words, we discard small coefficients. 

Beside the mentioned approach to exploit sparsity, we also use the  
set-membership filtering (SMF) technique \cite{yazdanpanah2017robustness,Diniz_adaptiveFiltering_book2013,Zheng7892916,Bhotto6217758,Yazdanpanah7456244,Liu_8320072,sharafi2020robustness} to propose  
the Low-Complexity Set-Membership Normalized Least-Mean-Square (LCSM-NLMS1) algorithm. This algorithm combines the set-membership normalized least-mean-square (SM-NLMS) algorithm~\cite{Zhang6823156} 
with the discarding technique to exploit sparsity. 
Indeed, the SMF strategy avoids new update when the error is greater than the pre-determined upper bound; thus, it can decrease the computational load even further. 

This paper is organized as follows.
In Section~\ref{sec:SMF}, we review the SMF technique. 
The LCSM-NLMS1 and the LCSM-NLMS2 algorithms are proposed in Section~\ref{sec:lcsm-nlms}. 
Numerical results and conclusions are described in Sections~\ref{sec:simulations} and~\ref{sec:conclusions}, respectively.


\section{Set-Membership Filtering (SMF)} \label{sec:SMF}

In the last decades, the SMF is proposed to acquire the adaptive coefficients $\wbf$ so that the 
magnitude of their error is upper bounded by a constant positive factor 
$\gammabar \in \mathbb{R_+}$. 
For appropriately adopted $\gammabar$, we can have various acceptable estimates for $\wbf$. 
Assume that ${\cal S}$ is the set of all input-desired data $(\xbf,d)$, and define the feasibility set $\Theta$ as
\begin{align}
\Theta=\bigcap_{(\xbf,d)\in{\cal S}}\{\wbf\in\mathbb{R}^{N+1}:|d-\wbf^T\xbf|\leq\gammabar\}.
\end{align}
Moreover, define the constraint set ${\cal H}(k)$ containing all vectors $\wbf$ so that, at a given iteration $k$, the magnitude of their errors are upper bounded by $\gammabar$, that is
\begin{align}
{\cal H}(k)=\{\wbf\in\mathbb{R}^{N+1}:|d(k)-\wbf^T\xbf(k)|\leq\gammabar\},
\end{align}
where $\wbf$, $\xbf(k)$, and $d(k)$ are the weight vector, the input vector, and 
the desired signal, respectively. 
The membership set $\psi(k)$ can be described by
\begin{align}
\psi(k)=\bigcap_{i=0}^k{\cal H}(i).\label{eq:set-membership_set}
\end{align}
Note that, as $k \rightarrow \infty$, the membership set converges to the feasibility set.
However, we cannot compute $\psi(k)$ because of practical issues; thus, we calculate a point estimate by the information obtained by the constraint set ${\cal H}(k)$~\cite{Zhang6823156}.

 




\section{{The Low-Complexity Set-Membership NLMS Algorithm}} \label{sec:lcsm-nlms}

To exploit sparsity in unknown systems with a computational complexity lower than that of the existing sparsity-aware algorithms, we propose the Low-Complexity Set-Membership NLMS (LCSM-NLMS) algorithm.
To this end, we avoid updating the small adaptive coefficients of the sparse system.
Thus, first, in Subsection~\ref{sub:LCSM-NLMS1}, we propose the LCSM-NLMS1. 
Then, we discuss some important issues of the LCSM-NLMS1 algorithm.
Moreover, in Subsection~\ref{sub:LCSM_NLMS2}, we introduce the LCSM-NLMS2 algorithm (an improved version of the LCSM-NLMS1) intending to reduce computational cost even further.


\subsection{LCSM-NLMS1 algorithm} \label{sub:LCSM-NLMS1}

To exploit systems sparsity by the LCSM-NLMS1 algorithm, the thresholding approach by the discard function is applied. The discard function, $f_\epsilon:\mathbb{R}\rightarrow\mathbb{R}$, is defined by~\cite{yazdanpanah2019improved,Yazdanpanah7760558}
\begin{align}
f_\epsilon(w)=\left\{\begin{array}{ll}w&{\rm if~} |w|\geq \epsilon  \\0&|w|< \epsilon \end{array}\right.,\label{eq:f_epsilon}
\end{align}
where $\epsilon$ is a positive constant.
In fact, this function disregards those entries of $w$ that are close to zero. 
Note that $\epsilon$ determines which values are close to zero, and some prior knowledge about the sparse system can be used to select this parameter.
It is worth mentioning that $f_\epsilon(w)$ is not differentiable at $\pm \epsilon$, and we need its derivative in the optimization problem.
To resolve this problem, we assume that the derivative of $f_\epsilon(w)$ at $+\epsilon$ and $-\epsilon$ is zero. 
We can know introduce the discard vector function $\fbf_\epsilon:\mathbb{R}^{N+1}\rightarrow\mathbb{R}^{N+1}$ by
\begin{align}
\fbf_\epsilon(\wbf)=[f_\epsilon(w_0),\cdots,f_\epsilon(w_N)]^T.
\end{align}

When the output estimation error is greater than the predetermined positive value $\gammabar$, the LCSM-NLMS1 algorithm updates the adaptive coefficients whose absolute values are greater than $\epsilon$. 
Whenever $\wbf(k)\not\in{\cal H}(k)$, i.e., $|e(k)|=|d(k)-\wbf^T(k)\xbf(k)|>\gammabar$, the optimization criterion of the LCSM-NLMS1 is given by
\begin{align}
&\min \frac{1}{2}\|\fbf_\epsilon(\wbf(k+1))-\wbf(k)\|^2 \nonumber\\
&{\rm subject~to}\nonumber\\
&d(k)-\wbf^T(k+1)\xbf(k)=\gammabar.\label{eq:ssm_optimization}
\end{align}

To compute the solution of this optimization problem, the Lagrangian $\mathbb{L}$ should be formed as
\begin{align}
\mathbb{L}=&\frac{1}{2}\|\fbf_\epsilon(\wbf(k+1))-\wbf(k)\|^2+\lambda(k)[d(k)-\wbf^T(k+1)\xbf(k)-{\gammabar}],
\end{align}
where $\lambda(k)\in\mathbb{R}$ is the Lagrange multiplier. 
By taking the differential of the above equation with respect to $\wbf(k+1)$ and letting it equal to zero, we get
\begin{align}
\fbf_\epsilon(\wbf(k+1))=\wbf(k)+\lambda(k)\Fbf_\epsilon^{-1}(\wbf(k+1))\xbf(k),\label{eq:F_a(w(k+1))}
\end{align}
where $\Fbf_\epsilon(\wbf(k+1))$ is the Jacobian matrix of $\fbf_\epsilon(\wbf(k+1))$.
To form the recursion, we use the projection approximation subspace tracking with deflation approach, as in~\cite{Wang_WirelessCommunicationSystems_book2004}, in~\eqref{eq:F_a(w(k+1))}, then we can substitute $\fbf_\epsilon(\wbf(k+1))$ and $\Fbf_\epsilon^{-1}(\wbf(k+1))$ with $\wbf(k+1)$ and $\Fbf_\epsilon^{-1}(\wbf(k))$, respectively. 
Thus, we attain
\begin{align}
\wbf(k+1)=\wbf(k)+\lambda(k)\Fbf_\epsilon^{-1}(\wbf(k))\xbf(k).\label{eq:w(k+1)-ssm}
\end{align}
By replacing this equation into the constraint relation \eqref{eq:ssm_optimization}, we obtain $\lambda(k)$ as
\begin{align}
\lambda(k)=\frac{e(k)-\gammabar}{\xbf^T(k)\Fbf_\epsilon^{-1}(\wbf(k))\xbf(k)}.\label{eq:lambda-ssm}
\end{align}
Then, by substituting (\ref{eq:lambda-ssm}) into (\ref{eq:w(k+1)-ssm}), we get the following recursion rule 
\begin{align}
\wbf(k+1)=\wbf(k)+\frac{(e(k)-\gammabar)\Fbf_\epsilon^{-1}(\wbf(k))\xbf(k)}{\xbf^T(k)\Fbf_\epsilon^{-1}(\wbf(k))\xbf(k)+\delta},
\end{align}
where $\delta$ is a small positive constant to avoid division by zero.
Note that $\Fbf_\epsilon(\wbf(k))$ is a singular matrix; thus, the Moore-Penrose pseudoinverse (generalization of the inverse matrix) can be employed in the place of the standard matrix inversion. 
But $\Fbf_\epsilon(\wbf(k))$ is a diagonal matrix whose diagonal components are zero or one. 
In fact, when a coefficient of $\wbf(k)$ has the absolute value greater than $\epsilon$, then its corresponding entry on the diagonal of $\Fbf_\epsilon(\wbf(k))$ is one, but the remaining entries are zero. 
Note that the pseudoinverse of $\Fbf_\epsilon(\wbf(k))$ is $\Fbf_\epsilon(\wbf(k))$. 
Hence, the update rule is given by
\begin{align}
\wbf(k+1)=\wbf(k)+\frac{(e(k)-\gammabar)\Fbf_\epsilon^{-1}(\wbf(k))\xbf(k)}{\xbf^T(k)\Fbf_\epsilon^{-1}(\wbf(k))\xbf(k)+\delta}. \label{eq:update_equation}
\end{align} 
By changing $\gammabar$ to $\frac{\gammabar e(k)}{|e(k)|}$, we obtain the update equation of the LCSM-NLMS1 algorithm as follows
\begin{align}
\wbf(k+1)=\wbf(k)+\mu(k)\frac{e(k)\Fbf_\epsilon^{-1}(\wbf(k))\xbf(k)}{\xbf^T(k)\Fbf_\epsilon^{-1}(\wbf(k))\xbf(k)+\delta},
\end{align}
where
\begin{align}
\mu(k)=\left\{\begin{array}{ll}1-\frac{\gammabar}{|e(k)|}&\text{if~}|e(k)|>\gammabar,\\0&\text{otherwise.}\end{array}\right. \label{eq:mu(k)}
\end{align}

\subsection{Some discussion of the LCSM-NLMS1 algorithm}

The recursion rules of the LCSM-NLMS1 and SM-NLMS algorithms are similar, but the LCSM-NLMS1 algorithm only updates the subset of coefficients of $\wbf(k)$ that their absolute values are greater than $\epsilon$. 
Therefore, the LCSM-NLMS1 algorithm has lower computational cost in comparison with the SM-NLMS algorithm.

\begin{table*}[t]
	\begin{center}
		\caption{Number of required real multiplications and divisions for the SM-PNLMS, SM-$l_0$-NLMS, and LCSM-NLMS1 algorithms \label{tab1}}
		\begin{tabular}{|l|c|c|c|} \hline
			Algorithm & Addition $\&$ Subtraction & Multiplication & Division\\\hline
			SM-PNLMS & $N^2+5N+5$ & $7N+8$ & $2N+4$\\\hline
			SM-$l_0$-NLMS & $7N+7$ & $9N+11$ & $N+3$ \\\hline
			LCSM-NLMS1 & $3N+4$ & $3N+4$ & 1 \\\hline
		\end{tabular}
	\end{center}
\label{tb:computation}
\end{table*}

In Table~\ref{tb:computation}, for each update, we describe the computational complexity of the set-membership proportionate NLMS 
(SM-PNLMS)~\cite{Diniz_sm_pap_jasmp2007}, the set-membership $l_0$-NLMS (SM-$l_0$-NLMS)~\cite{Markus_sparseSMAP_tsp2014}, and the LCSM-NLMS1 algorithms. 
It is worth mentioning that, in Table~\ref{tb:computation}, the number of multiplications and additions are described for the update of all coefficients.
In means that we presented the worst situation for the LCSM-NLMS1 algorithm (i.e., $\epsilon=0$).
In this case, the computational load of the LCSM-NLMS1 algorithm is identical to that of the SM-NLMS algorithm. 
However, in practice $\epsilon$ is different from zero. Also, note that the number of divisions in the LCSM-NLMS1 is only one, while the SM-PNLMS and the SM-$l_0$-NLMS algorithms require $2N+4$ and $N+3$ divisions, respectively.
Furthermore, note that the memory requirements of the LCSM-NLMS1 algorithm are exactly the same as the NLMS algorithm.

Moreover, we must remind that the weight vector of the LCSM-NLMS1 algorithm cannot be initialized with the zero vector.
As the matter of fact, for this algorithm, the adaptive coefficients must be initialized by some values outside the interval $[-\epsilon,\epsilon]$; i.e., $|w_i (0)| > \epsilon$ for $i=0,1,\cdots,N$.

\subsection{The LCSM-NLMS2 algorithm} \label{sub:LCSM_NLMS2}

In the recursion rule of the LCSM-NLMS1, we can observe that for those coefficients of the adaptive filter fall inside $[-\epsilon,+\epsilon]$, the algorithm does note update them in the subsequent iterations since they are disregarded by the discard function.
Furthermore, the coefficients inside $[-\epsilon,+\epsilon]$ are close to zero and the they can be estimated by zero.
Also, inserting zero for these coefficients leads to a reduction of computational cost in computing the output signal $y(k) = \xbf^T(k) \wbf(k)$.
To this end, we introduce the LCSM-NLMS2 algorithm by multiplying $\wbf(k)$ by $\Fbf_\epsilon(\wbf(k))$.
Thus the update equation of the LCSM-NLMS2 is given by
\begin{align}
\wbf(k+1)=\Fbf_\epsilon(\wbf(k))\wbf(k)+\mu(k)\frac{e(k)\Fbf_\epsilon^{-1}(\wbf(k))\xbf(k)}{\xbf^T(k)\Fbf_\epsilon^{-1}(\wbf(k))\xbf(k)+\delta},
\end{align}
where $\mu(k)$ is described by~\eqref{eq:mu(k)}.


\section{Simulations} \label{sec:simulations}

In this section, we use the SM-PNLMS, theSM-$l_0$-NLMS, and the LCSM-NLMS2 algorithms, in system identification scenarios, to identify three sparse systems of order 12.
The impulse response of the sparse systems are presented in Table~\ref{tb:systems}.
We only presents the performance of the LCSM-NLMS2 algorithms since it requires lower computational load and performs better than LCSM-NLMS1 algorithm in sparse domains.
The input signal is a zero-mean white Gaussian noise with unit variance.
The additive noise has a zero-mean white Gaussian distribution with variance $\sigma_n^2=0.01$.
The threshold parameter $\gammabar$ is chosen as $\sqrt{5\sigma_n^2}$.
All algorithms are initialized with $\wbf(0)=0.1\times[1,\cdots,1]^T$, and the regularization parameter is adopted as $\delta=10^{-12}$.
For the LCSM-NLMS2 algorithm, $\epsilon$ is selected as $0.0001$.
For the SM-PNLMS and the SM-$l_0$-NLMS algorithms, the parameters $\alpha$, $\beta$, and $\varepsilon$ are adopted as $0.005$, $5$, and $100$, respectively.
The learning curves are obtained by averaging the outcomes of 500 runs. 

\begin{table*}
	\begin{center}
		\caption{The coefficients of unknown systems System 1, System 2, and System 3 \label{tab2}}
		\begin{tabular}{|c|ccccccccccccc|}\hline
			System 1&{\bf 0.02}&0&0&0&0&{\bf 0.6}&0&0&{\bf 0.25}&0&0&0&0\\\hline
			System 2&0&0&0&0&{\bf 0.3}&{\bf 0.6}&{\bf -0.5}&{\bf 0.7}&0&0&0&0&0\\\hline
			System 3&0&0&0&0&{\bf 0.3}&{\bf 0.5}&{\bf 0.7}&{\bf 0.5}&{\bf 0.3}&0&0&0&0\\\hline
		\end{tabular}
	\end{center}
\label{tb:systems}
\end{table*}

\begin{figure}[t!]
	\centering
	\includegraphics[width=0.75\linewidth]{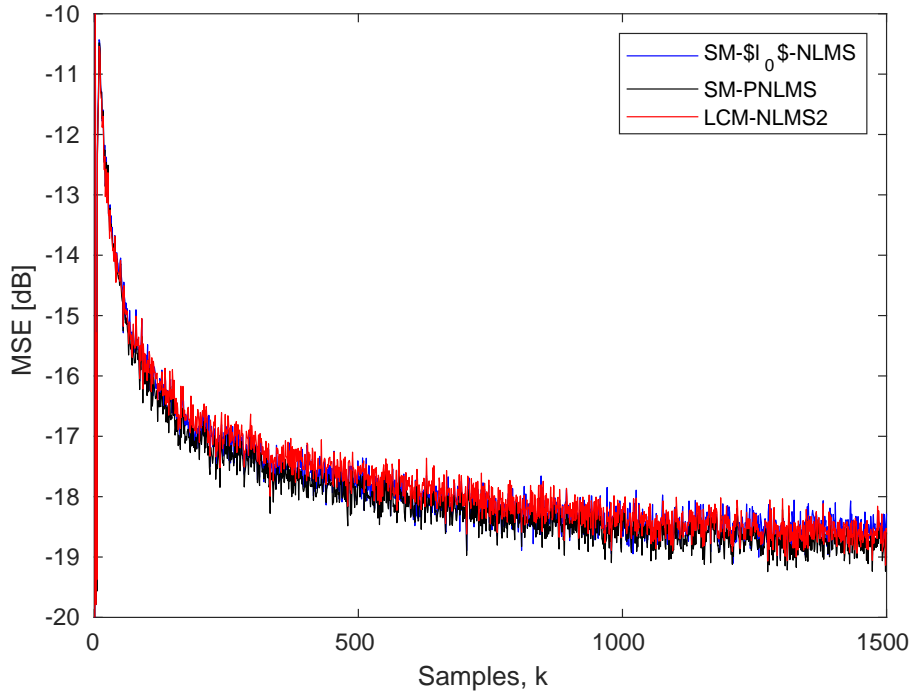}
	\caption{The MSE learning curves of the SM-PNLMS, the SM-$l_0$-NLMS, and the LCSM-NLMS2 algorithms when the unknown system is System 1. \label{fig:system1}}
\end{figure}

Figure~\ref{fig:system1} shows the MSE learning curves of the SM-PNLMS, the SM-$l_0$-NLMS, and the LCSM-NLMS2 algorithms when the system 1 is used as the unknown system.
We can see that the LCSM-NLMS2 algorithm has competitive performance with other two algorithms. 
In other words, the LCSM-NLMS2 has extremely similar performance to the SM-$l_0$-NLMS and the SM-PNLMS algorithms, whereas it requires very lower computational loads. 
Indeed, at each iteration during the steady-state, whenever an update is implemented, it updates only three coefficients, i.e., the coefficients greater than $0.0001$.
Moreover, the update rates of the SM-PNLMS, the SM-$l_0$-NLMS, and the LCSM-NLMS2 algorithms are 12.43$\%$, 20.06$\%$, and 13.35$\%$, respectively.
Therefore, the LCSM-NLMS2 algorithm can attain lower update rate as well.

\begin{figure}[t!]
	\centering
	\includegraphics[width=0.75\linewidth]{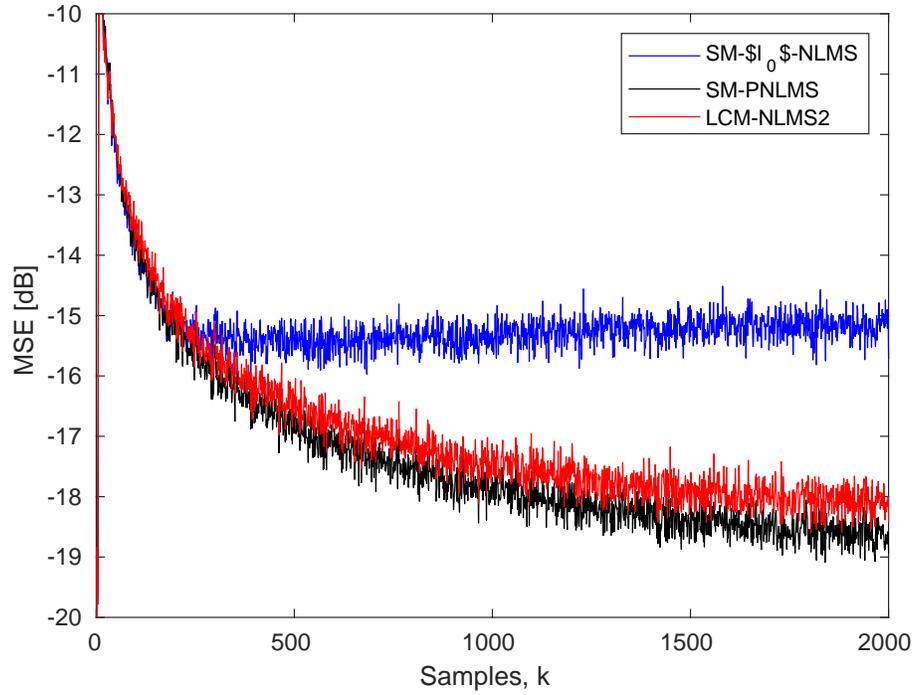}
	\caption{The MSE learning curves of the SM-PNLMS, the SM-$l_0$-NLMS, and the LCSM-NLMS2 algorithms when the unknown system is System 2. \label{fig:system2}}
\end{figure}

Figure~\ref{fig:system2} depicts the MSE learning curves of the SM-PNLMS, the SM-$l_0$-NLMS, and the LCSM-NLMS2 algorithms when the system 2 is considered as the unknown system.
In the case of block sparse system, we can observe that the LCSM-NLMS2 can attain similar steaty-state MSE to the SM-PNLMS algorithm; however, the SM-$l_0$-NLMS algorithm has remarkably higher steady-state MSE.
Also, in the steady-state, the LCSM-NLMS2 algorithm updates only four coefficients, when as update is executed. 
Furthermore, the update rate of the SM-PNLMS, the SM-$l_0$-NLMS, and the LCSM-NLMS2 algorithms are 10.70$\%$, 19.93$\%$, and 11.92$\%$, respectively.

\begin{figure}[t!]
	\centering
	\includegraphics[width=0.75\linewidth]{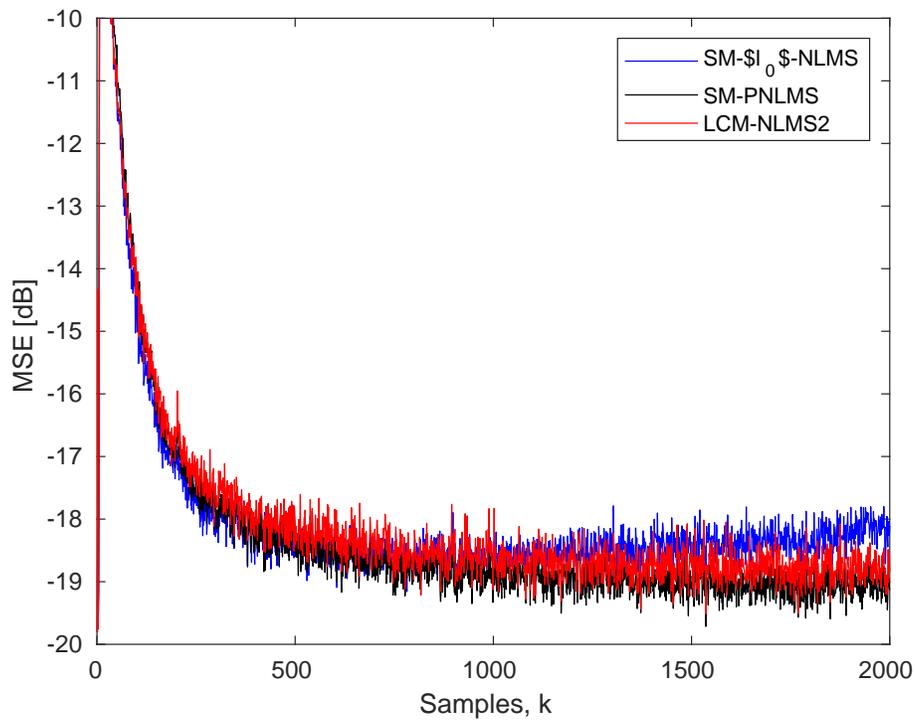}
	\caption{The MSE learning curves of the SM-PNLMS, the SM-$l_0$-NLMS, and the LCSM-NLMS2 algorithms when the unknown system is System 3. \label{fig:system3}}
\end{figure}

Figure~\ref{fig:system3} illustrates the MSE learning curves of the SM-PNLMS, the SM-$l_0$-NLMS, and the LCSM-NLMS2 algorithms when the system 3 is considered as the unknown system 3.
In the case of symmetric block sparse system, we can see that the SM-PNLMS and the LCSM-NLMS2 algorithms have similar performance; however, the LCSM-NLMS2 requires extremely lower computational resources.
Also, note that the MSE of the SM-$l_0$-NLMS algorithm started increasing after the iteration 1000.
Moreover, the update rates of the SM-PNLMS, the SM-$l_0$-NLMS, and the LCSM-NLMS2 algorithms are 7.65$\%$, 8.33$\%$, and 8.04$\%$, respectively.
Therefore, the LCSM-NLMS2 attains the lower computational cost since it updates only five coefficients during the steady-state, whenever an update is implemented.

\section{Conclusions} \label{sec:conclusions}

In this paper, the LCSM-NLMS1 and the LCSM-NLMS2 algorithms have been introduced to take benefit of sparsity in the
signal systems and to reduce computational cost.
For this purpose, a simple update equation has been derived so that it only updates the coefficients whose magnitudes are greater than a pre-defined positive value.
Moreover, this approach is used with the set-membership technique to attain even lower computational burden and and low update rate.
The numerical results have presented the superior performance of the LCSM-NLMS2 algorithm to some other sparsity-aware set-membership adaptive filters regarding the computational resources.
In other words, the LCSM-NLMS2 algorithm executed as well as the SM-PNLMS and the SM-$l_0$-NLMS algorithms, whereas requiring fewer arithmetic operations..


%



%
%

\ifCLASSOPTIONcaptionsoff
  \newpage
\fi

\normalsize
\bibliographystyle{IEEEbib}
\bibliography{My_bib}

\end{document}